\documentclass[final,5p,times,twocolumn]{elsarticle} 

\usepackage{lineno,hyperref}

\journal{Nuclear Instruments and Methods in Physics Research A}
\begin{document}

\begin{frontmatter}

\title{Development of
Si-CMOS hybrid detectors towards
electron tracking based Compton imaging in semiconductor detectors}

\author[Tokyo,ISAS]{Hiroki Yoneda\corref{correspondingauthor}}
\cortext[correspondingauthor]{Corresponding author}
\ead{yoneda@astro.isas.jaxa.jp}

\author[Rikyo]{Shinya Saito}

\author[ISAS,Tokyo]{Shin Watanabe}

\author[ISAS]{Hirokazu Ikeda}

\author[ISAS,Tokyo]{Tadayuki Takahashi}

\address[Tokyo]{University of Tokyo, 7-3-1 Hongo, Bunkyo, Tokyo 113-0033, Japan}
\address[ISAS]{The Institute of Space and Astronautical Science, Japan Aerospace Exploration Agency (ISAS/JAXA), 3-1-1 Yoshinodai Chuo-ku Sagamihara Kanagawa 252-5210, Japan}
\address[Rikyo]{Rikkyo University, 3-34-1 Nishi Ikebukuro, Toshima-ku, Tokyo 171-8501, Japan}

\begin{abstract}
Electron tracking based Compton imaging is a key technique to improve the sensitivity of Compton cameras
by measuring the initial direction of recoiled electrons.
To realize this technique in semiconductor Compton cameras, we propose a new detector concept, Si-CMOS hybrid detector.
It is a Si detector bump-bonded to a CMOS readout integrated circuit to obtain electron trajectory images.
To acquire the energy and the event timing, signals from N-side are also read out in this concept.
By using an ASIC for the N-side readout, the timing resolution of few $\mathrm{\mu s}$ is achieved.
In this paper, we present the results of two prototypes with 20 $\mathrm{\mu m}$ pitch pixels.
The images of the recoiled electron trajectories are obtained with them successfully.
The energy resolutions (FWHM) are $4.1~\mathrm{keV}$ (CMOS) and $1.4~\mathrm{keV}$ (N-side) at 59.5 keV.
In addition, we confirmed that the initial direction of the electron is determined using the reconstruction algorithm based on the graph theory approach.
These results show that Si-CMOS hybrid detectors can be used for electron tracking based Compton imaging.
\end{abstract}

\end{frontmatter}


\section{Introduction}
Compton imaging technique is one of the most powerful techniques for gamma-ray imaging.
Especially in the energy band from few 100 keV to few MeV,
it has been used in various fields including astrophysics\citep{COMPTEL:1993, SGD:2016}, medical imaging\citep{Takeda:2012} and radiation monitoring\citep{takahashi:2012, takeda:2015}
because Compton scattering becomes the dominant interaction of photons with matter in this energy range.
In Compton imaging, by measuring interaction positions and deposit energies of a Compton-scattered gamma-ray,
the scattering angle is calculated and the incoming direction of the gamma-ray is constrained on a cone based on the kinematics of Compton scattering.
The incoming direction is determined as the intersection of the Compton cones.
Thus we need a number of photons to identify a certain intersection as the incoming direction statistically significantly.
If a gamma-ray incoming direction is constrained on a smaller region than the Compton cone,
we can identify it with fewer photons.

To tackle this problem, the most promising solution is to measure the initial direction of the recoiled electron
from the first interaction of the scattering.
The direction of an in-coming gamma-ray is constrained on an arc-shape region based on the momentum conservation law.
This approach is realized in gaseous Compton cameras
but the detector size has to be large due to the low density, and the energy resolution is limited\citep{TAKADA:2005,KABUKI:2007}. 
On the other hand, in semiconductor Compton cameras which have advantages with the high density and the high energy resolution,
only the demonstration in the limited conditions is achieved\citep{Vetter:2011}.

\begin{figure}[!ht]
  \begin{center} 
    \includegraphics[bb = 0 0 567 229, width = 85 mm]{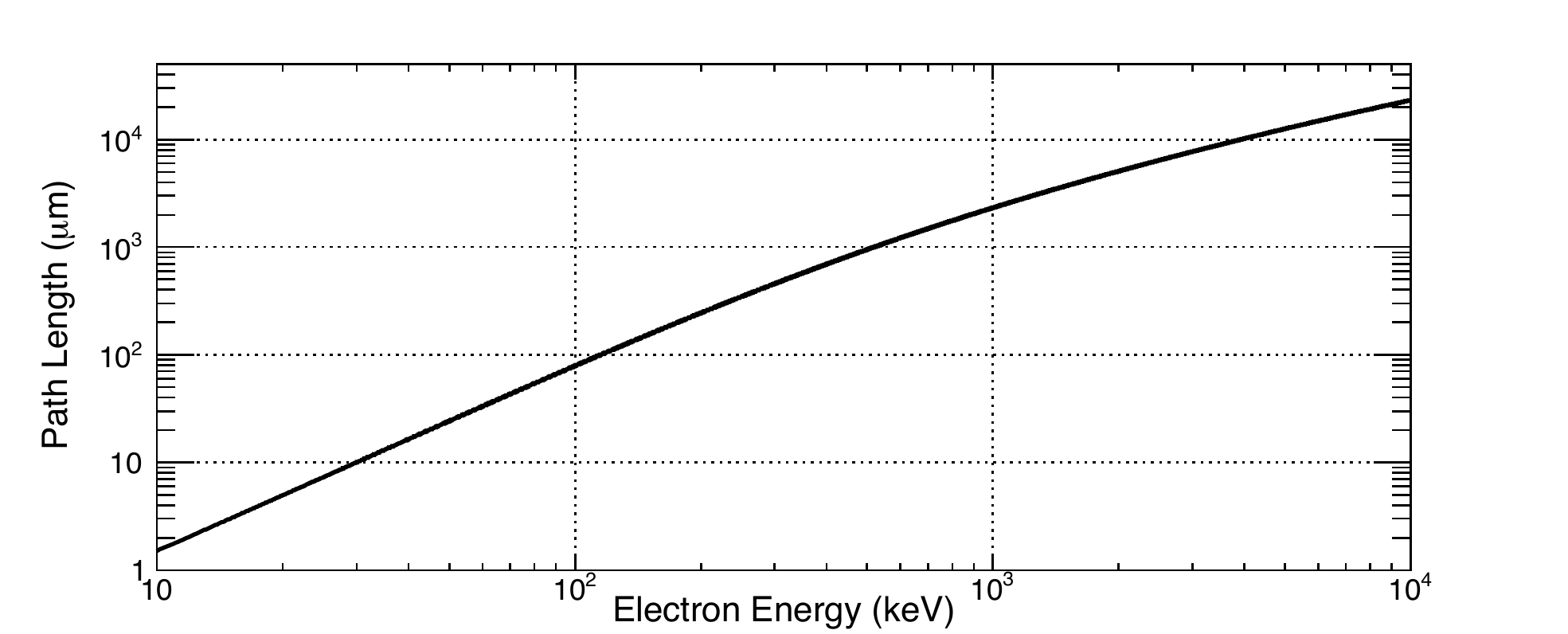}
    \caption{A curve between mean length and energy of electron trajectories in Si. The calculation is based on Bethe formula described in \citep{Knoll:2010}. Only collisional loss is considered because the electron energy is lower than few MeV and bremsstrahlung is negligible.} 
    \label{fig:path_length} 
  \end{center}
\end{figure}

To utilize the electron tracking based imaging in semiconductor Compton cameras,
the detector should meet several requirements.
Figure~\ref{fig:path_length} shows the mean length of electron trajectories in Si.
It is $\sim100~\mathrm{\mu m}$ for the electron of $\sim$ 100 keV.
($\sim 100~\mathrm{keV}$ is transferred to the recoiled electron when a 500 keV gamma-ray is scattered at an angle of 45$^\circ$.)
Because the direction of an electron momentum changes along its trajectory due to multiple scattering,
the first part of the trajectory is important.
To resolve the trajectory of $\sim100~\mathrm{\mu m}$, $\sim10~\mathrm{\mu m}$ spatial resolution is required.
The second requirement is fast timing resolution to identify coincidence events in scatter and absorber detectors.
A trial of the electron tracking based Compton imaging, by combining a scientific CCD and a Ge detector, was reported by Vetter et. al.\citep{Vetter:2011}.
In the experiment, the CCD provided the time of a frame with the frame rate of $\sim$ 0.5 fps.
The coincidence events between the CCD and the Ge detector were determined only by the Compton kinematics ({\it i.e.} the relationship of measured energies and positions on detectors).
In order to select a clean event which includes one electron trajectory in the frame, the event rate was limited below $\sim$ 2 Hz.
In order to use the camera in a practical environment, 
we need a fast event trigger of less than $\sim 10~\mathrm{\mu s}$ resolution from the device in conjunction with tracking information.
In addition, the thickness of the detector should be $\sim 500~\mathrm{\mu m}$
in terms of the efficiency and the electron trajectory length.
We can consider to use a fast readout CMOS pixel sensor\citep{MAPS:2009, Ishikawa:2017} alternatively 
to achieve both high spatial resolution and high timing resolution.
However its thickness is less than $\sim 20~\mathrm{\mu m}$ and electrons can escape from the detector easily in addition to the low efficiency.
Finally, for good spectral and spatial resolution of Compton imaging, high energy resolution is also required.
The energy resolution of a few percent is sufficient because the imaging performance is limited by Doppler broadening\citep{Zoglauer:2003}.

\label{sec:si-cmos}


\section{Si-CMOS hybrid detector}
\begin{figure}[!ht]
  \begin{center}
    \includegraphics[bb = 0 0 1000 600, width = 60 mm]{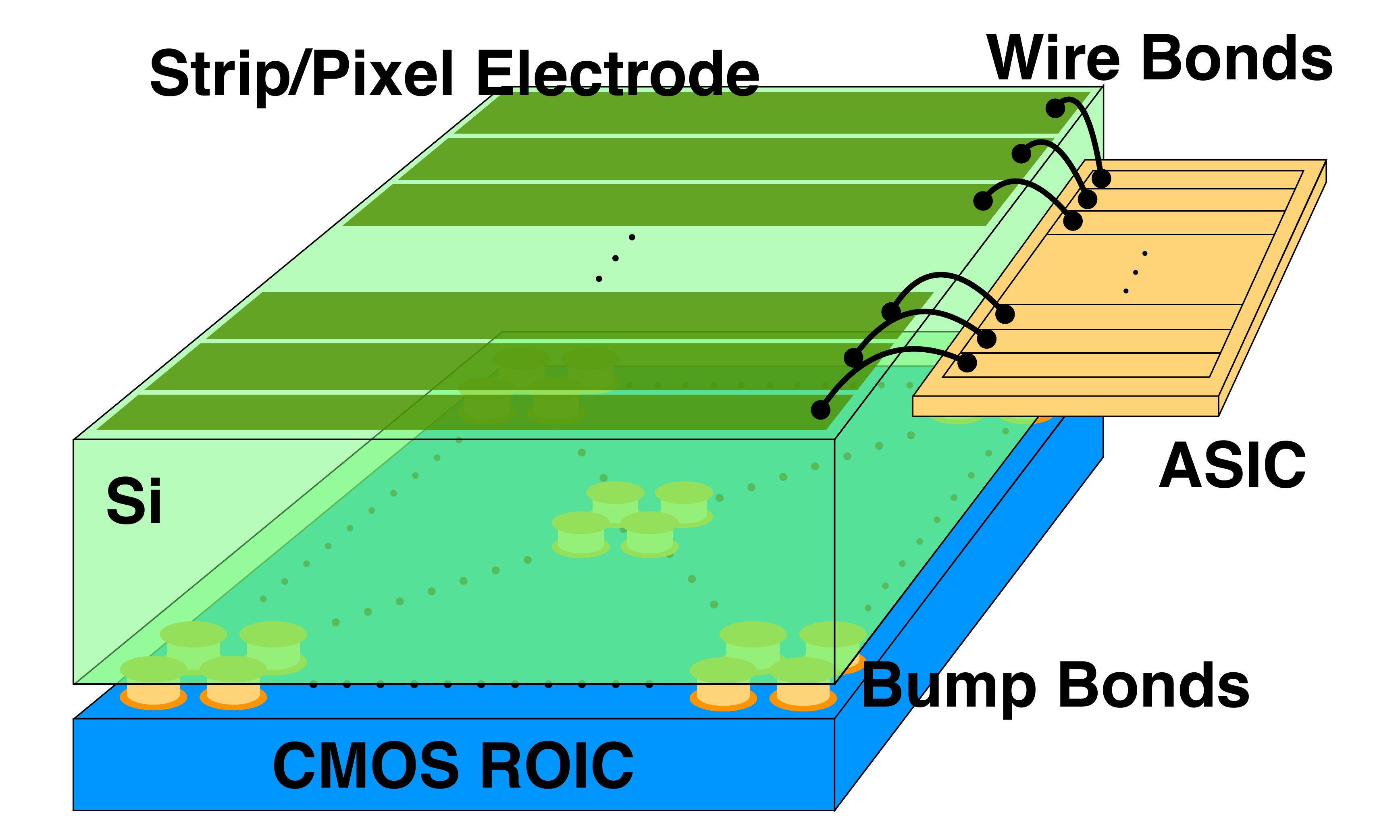}        
    \caption{A schematic view of Si-CMOS hybrid detector.} 
    \label{fig:concept} 
  \end{center}
  \end{figure}

In collaboration with Hamamatsu Photonics,
we have been developing Si-CMOS hybrid detector which is a Si detector bump-bonded to a CMOS readout integrated circuit (ROIC)
with 20 $\mathrm{\mu m}$ pitch pixels.
Figure~\ref{fig:concept} describes a schematic view of the detector.
Each image from the CMOS ROIC is read out non-destructively thus the CMOS ROIC is reset 
after taking a certain number of images from the last reset.
Negatively doped silicon strips or pixels (N-side) are implanted 
on the opposite side of the n-type silicon bulk.
Signals from these electrodes are also read out to measure the electron energy when one of their pulse heights becomes higher than a threshold using a low noise analog Application Specific Integrated Circuit (ASIC).
The ASIC is developed for the Soft Gamma-ray Detector on the {\it Hitomi} satellite\citep{SGD:2016}
and the event timing is also acquired with the resolution of few $\mathrm{\mu s}$.
A circuit diagram of the ASIC is shown in Figure 3 of \citep{SGD:2016}.
When an event trigger is generated on the ASIC,
information about the status of the CMOS ROIC (which row/image is being read out) is saved.
Using the energy and spatial and timing information from the ASIC,
the electron trajectories are identified with the ASIC events.

\begin{figure}[!htb]
  \begin{center}
    \includegraphics[bb = 0 0 1000 470, width = 75 mm]{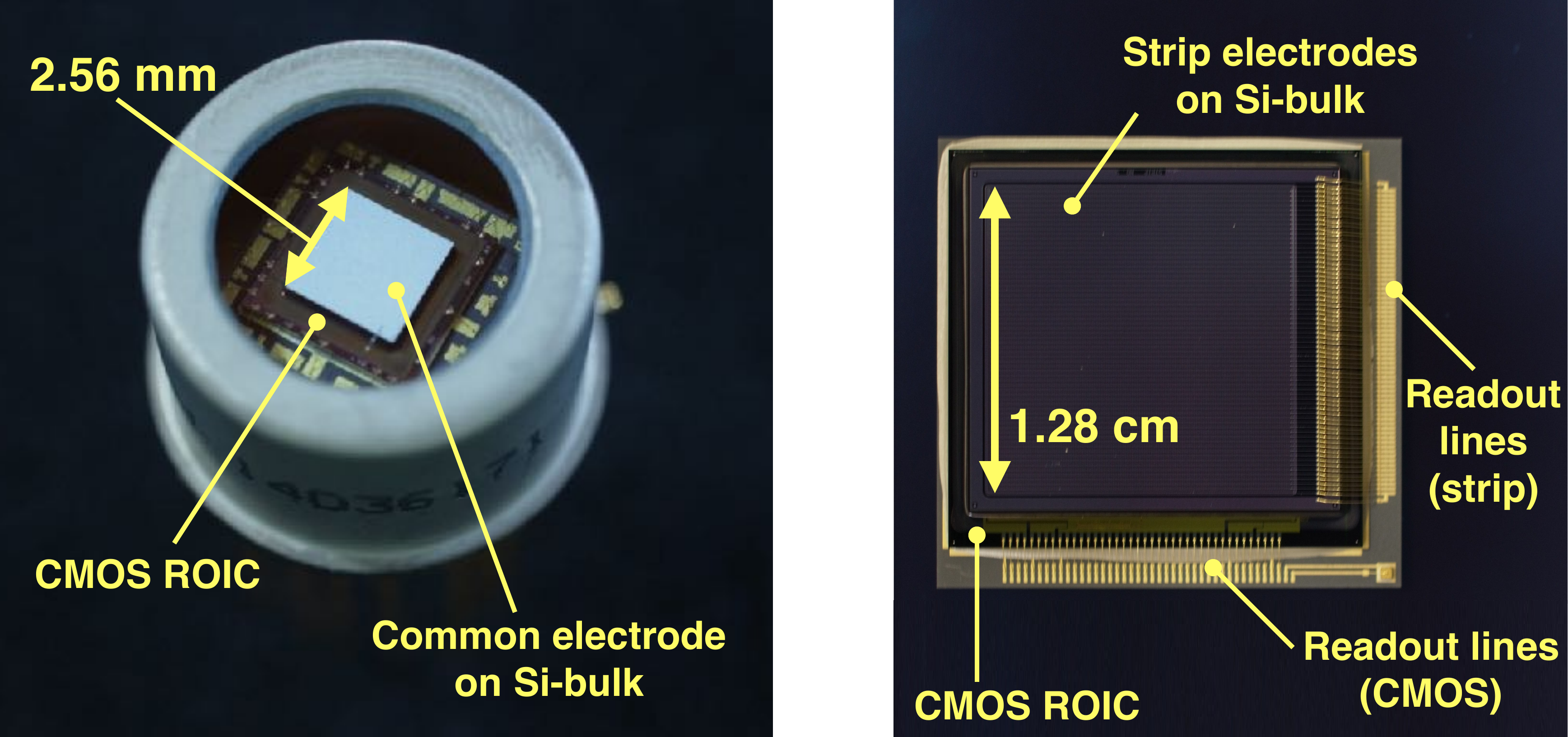}
  \end{center}
        \caption{Photos of the Si-CMOS hybrid detectors. Gamma-ray sources are located above the detectors at a distance of few cm in the experiments. (left) The first prototype with $128\times128$ pixels. (right) The second prototype with $640\times640$ pixels.} 
    \label{fig:prototype} 
\end{figure}

We developed two prototypes of Si-CMOS hybrid detector.
The left one in Figure~\ref{fig:prototype} is the first prototype.
The number of pixels is $128\times128$
and the detector size is $2.56~\mathrm{mm} \times 2.56~\mathrm{mm}$ with a thickness of $0.5~\mathrm{mm}$.
In this prototype, we focused on testing the bump-bonding between 
the Si bulk and the CMOS ROIC and evaluating the performance of the CMOS ROIC.
On N-side, the electrode is not segmented and signals from this side are not read out.
The second prototype is the right one in Figure~\ref{fig:prototype}.
The number of pixels is increased to $640\times640$
and the detector size is $12.8~\mathrm{mm} \times 12.8~\mathrm{mm}$.
The thickness of Si is the same as the first prototype.
N-side has 64 strip electrodes wire-bonded to the ASIC.
The strip pitch is 200 $\mathrm{\mu m}$ and the length is 12.8 $\mathrm{mm}$.
The frame readout rate can be controlled by the frequency of an input clock.
In this work, the first prototype was operated with 5 MHz clock and the frame rate is $\sim$ 60 fps.
The second one was operated with 2 - 4 MHz clock and the frame rate is $\sim$ 2 - 4 fps.
Because the pixel number of this prototype is 25 times larger than that of the first one,
the frame rate becomes lower.
In the future development, we will attempt to increase the frame rate using established technologies for fast readout CMOS pixel sensors.
We operated these detectors under a bias voltage of 200 V and
a temperature of $-20~{}^\circ\mathrm{C}$ or $-10~{}^\circ\mathrm{C}$.


\section{Data Process \& Detector Performance}
\label{sec:performance}
\subsection{CMOS}
\begin{figure}[!th]
  \begin{center} 
    \includegraphics[bb = 0 0 600 750, width = 75 mm]{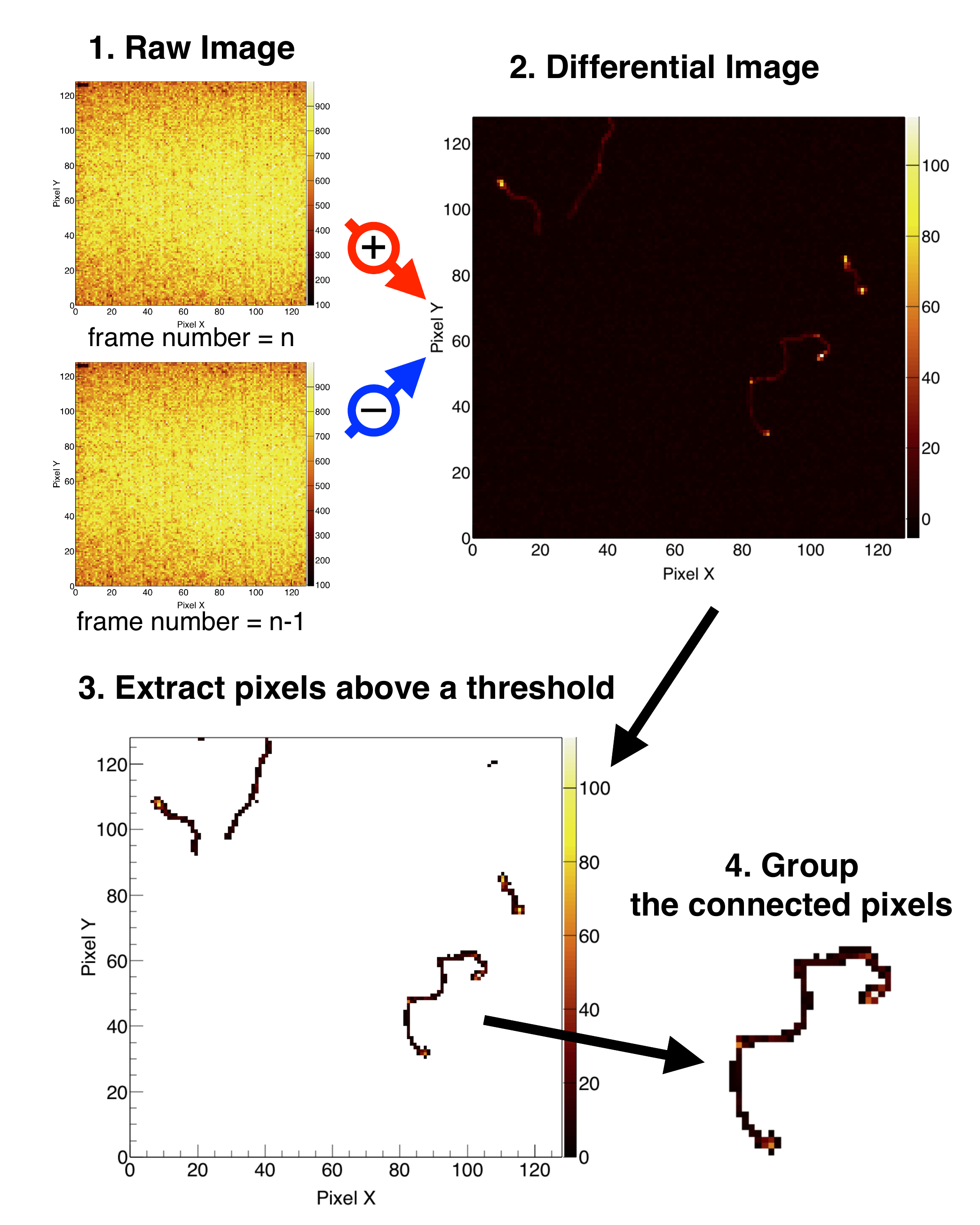}
    \caption{Data processing flow for our analysis of CMOS images} 
    \label{fig:cmos_dp} 
  \end{center}
\end{figure}
We analyzed the images from the CMOS ROIC using the first prototype
and evaluate its performance.
Figure~\ref{fig:cmos_dp} shows the data processing flow for our analysis.
Each pixel value in the raw image
corresponds to the accumulated charges in the pixel
from the last frame reset
because the CMOS ROIC is read out non-destructively.
To extract events in frame intervals,
we calculate the difference between frame images and their one before frame images.
Additionally, the reset noise (kTC noise) and the fixed pattern noise are canceled out in this process.
The median of the differential values in each row is calculated and subtracted from all the pixels in the row to estimate
an offset due to dark current and common-mode noise in the same row.
After these processes, differential images are produced as shown on the upper right in Figure~\ref{fig:cmos_dp}.
The electron trajectories on the images is produced due to Compton scattering or photoabsorption.
Thus when using the Si-CMOS detectors for Compton cameras, a coincidence setup is required to identify Compton scattering events.
However we do not need the setup here because we just test the images of the electrons.
The electron trajectories are not straight due to multiple scattering
and we need an appropriate algorithm to determine their initial direction as described in Section~\ref{sec:algorithm}.

\begin{figure}[!th]
  \begin{center} 
    \includegraphics[bb = 0 0 1230 400, width = 85 mm]{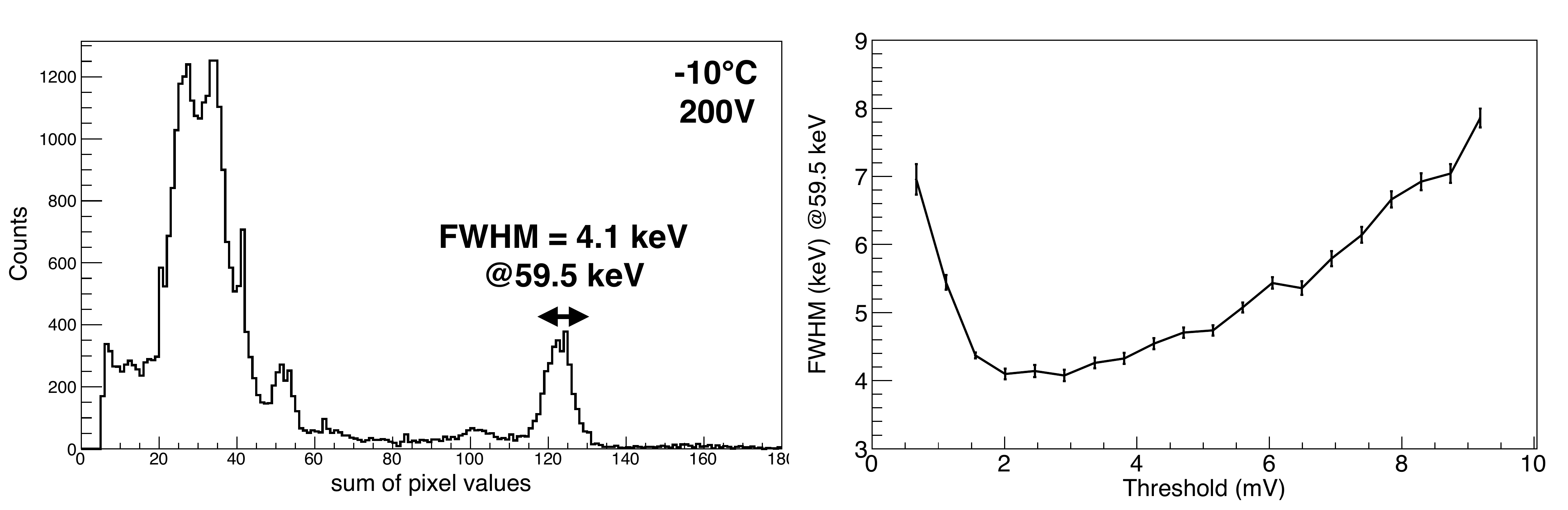}
    \caption{(left) CMOS energy spectrum with gamma-rays of $^{241}$Am. (right) Energy resolution (FWHM) depending on the threshold. 1 keV corresponds to $\sim$ 2 mV. In this analysis, we ignored the pixel groups consisting of only one pixel as the background.} 
    \label{fig:CMOS_spe} 
  \end{center}
\end{figure}

To separate the trajectories on the image,
we extract pixels above a threshold.
If an extracted pixel is included in adjacent pixels of another extracted pixel, 
the two pixels are regarded as "connected".
We make groups consisting of connected pixels which correspond to electron trajectories.
The electron energy is estimated by summing up the pixel values in one group.
Figure~\ref{fig:CMOS_spe} shows the energy spectrum of gamma-rays of $^{241}$Am.
The energy resolution depends on the threshold strongly (right in Figure~\ref{fig:CMOS_spe}).
When the threshold is set too low, many noise pixels are included in one pixel group.
On the other hand,
some pixels with real signals cannot be extracted
in the case that the threshold is too high.
When applied the optimal threshold, 
the energy resolution (FWHM) is $4.1~\mathrm{keV}$ (6.9\%) at 59.5 keV.

Around 59.5 keV, the distribution of the pixel number in the trajectory is described well with a Gaussian function.
In the energy range from 55 keV to 65 keV, its mean ($\mu$) is 5.7 pixels but
as shown in Figure~\ref{fig:path_length},
the mean length of the trajectories of 60 keV is theoretically $\sim 30~\mathrm{\mu m}$ larger than $\mu$.
This indicates that the image spreads over the true electron trajectory.
It is probably be caused by charge sharing due to interpixel capacitance\citep{PSU:2017}
and may affect the energy resolution.
However, using pixel values of adjacent pixels, we can reconstruct the electron trajectory well as described in Section~\ref{sec:algorithm}.
It might be better to enhance the charge sharing from the viewpoint of measuring the initial direction of the electron momentum.

\subsection{ASIC}
\begin{figure}[!hbt]
  \begin{center} 
    \includegraphics[bb = 0 0 567 409, width = 75 mm]{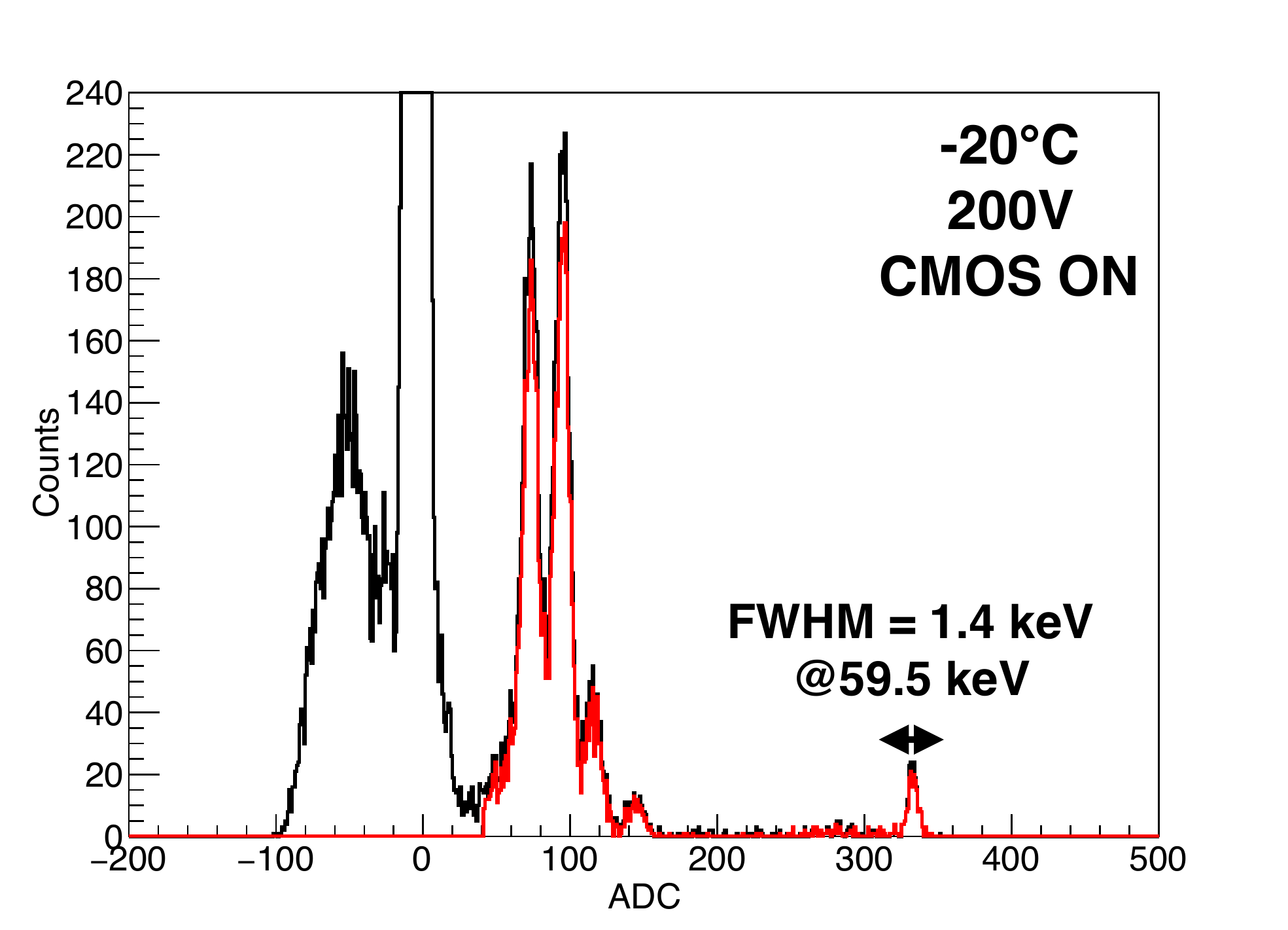}
    \caption{ASIC energy spectrum with gamma-rays of $^{241}$Am. The events before and after the row switching are ignored  (15\% of readout time). Red line is the spectrum with single-hit events.} 
    \label{fig:ASIC_spe} 
  \end{center}
\end{figure}

We also eveluate the spectral performance of the ASIC using the second prototype.
Figure~\ref{fig:ASIC_spe} is the energy spectrum with gamma-rays of $^{241}$Am in the operation under $-20~{}^\circ\mathrm{C}$.
The energy resolution (FWHM) is $1.4~\mathrm{keV}$ (2.3\%) at 59.5 keV when the CMOS ROIC and the ASIC are operated together.
We should mention that
an large noise is generated at N-side due to the interference with the CMOS ROIC in the current prototypes
when the CMOS ROIC switches the readout row.
The noise may be caused by an electric current following the CMOS ROIC switching or the surface of the CMOS ROIC which is not metal-coated, 
but it is currently under investigation.
In Figure~\ref{fig:ASIC_spe},
the events before and after the row switching (15\% of readout time) are ignored.
When the CMOS ROIC is turned off,
the energy resolution is improved to $1.1~\mathrm{keV}$ (1.8\%).

\begin{figure}[!htbp]
  \begin{center}
       \includegraphics[bb = 0 0 567 377, width = 75 mm]{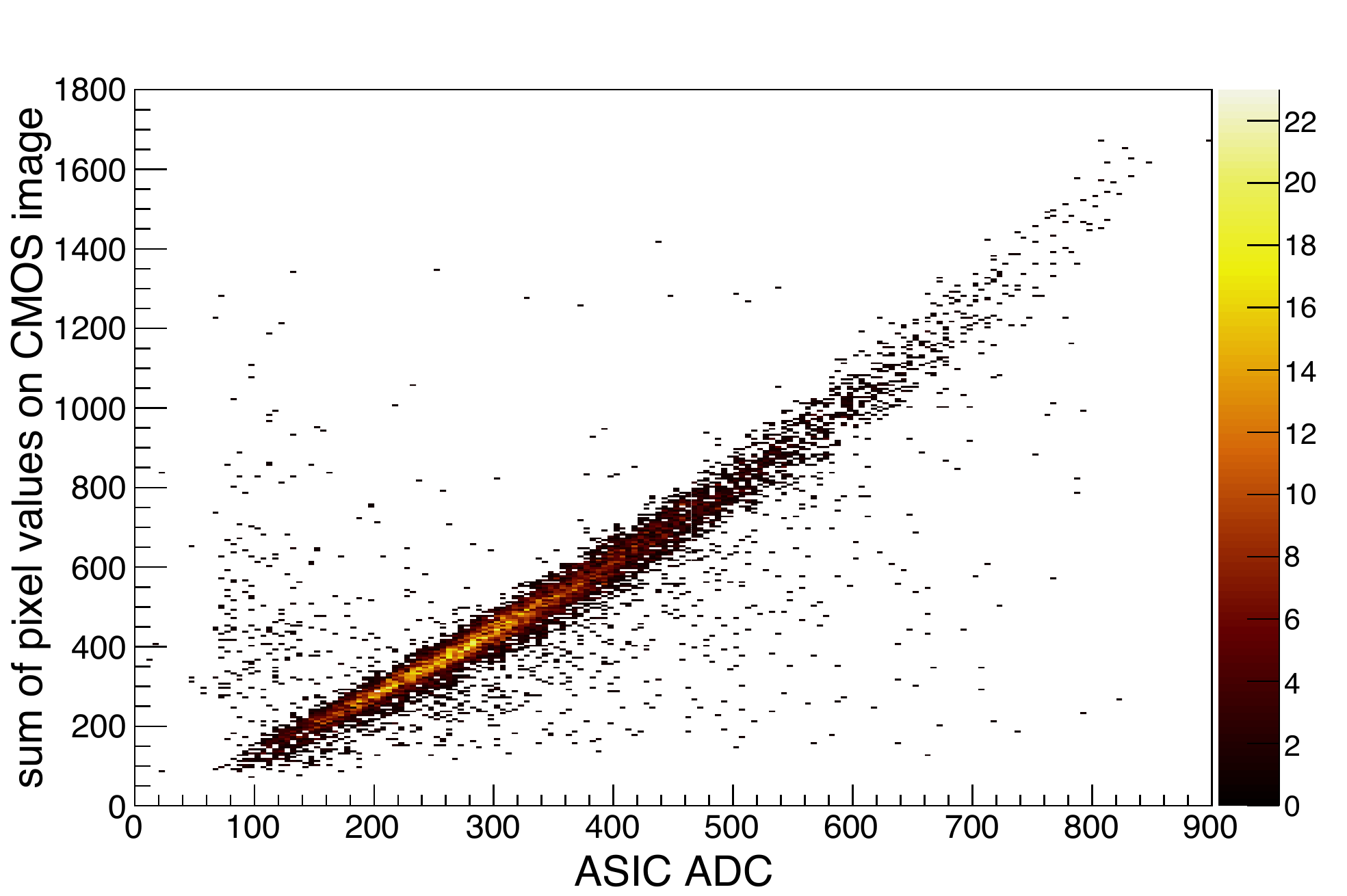} 
    \caption{Correlation between the sum of pixel values and ASIC ADC of the identified event pairs. $^{60}$Co is used for a gamma-ray source in this measurement.} 
    \label{fig:ASIC_corr} 
  \end{center}
\end{figure}

The position of an ASIC signal event is constrained in the strip electrode from which the signal is generated, and 
information on the CMOS ROIC (which row/image was being read out) is latched as soon as the trigger is generated in the ASIC. 
Using this information, we can identify ASIC events to trajectories on CMOS images.
When there is only one electron trajectory in a certain strip on the CMOS images,
and only one single-hit event exists on the strip in ASIC event lists in the corresponding frame interval, 
we select the pair of the ASIC event and the trajectory on the CMOS image.
Figure~\ref{fig:ASIC_corr} shows the correlation between the sum of pixel values and the ASIC ADC value of the event pairs.
The linear correlation in the energy is seen clearly. 
This indicates that the event identification works well.
Furthermore, using the energy information additionally, 
the event identification is possible even if several trajectories are on the same strip on the same image.

\section{Electron Trajectory Reconstruction Algorithm}
\label{sec:algorithm}

The direction of the initial electron momentum is the important physical parameter
in the electron tracking based Compton imaging.
The reconstruction algorithm to measure the initial direction has been investigated for gaseous detectors in X-ray polarimetry field.
In previous studies, the algorithm is often based on the moment analysis\citep{Bellazzini:2003, BLACK:2007}.
However, Li et al.\citep{Li:2017} pointed out that this can fail in some cases ({\it i.e.} the trajectory is a U-shape curve, and the initial and end position are close to each other.)
and they introduced the algorithm based on graph theory alternatively.
Here, we develop an algorithm to measure the initial electron direction using the CMOS images following their approach.
The details are described below.

\begin{figure}[!htb]
  \begin{center} 
    \includegraphics[bb = 0 0 700 350, width = 85 mm]{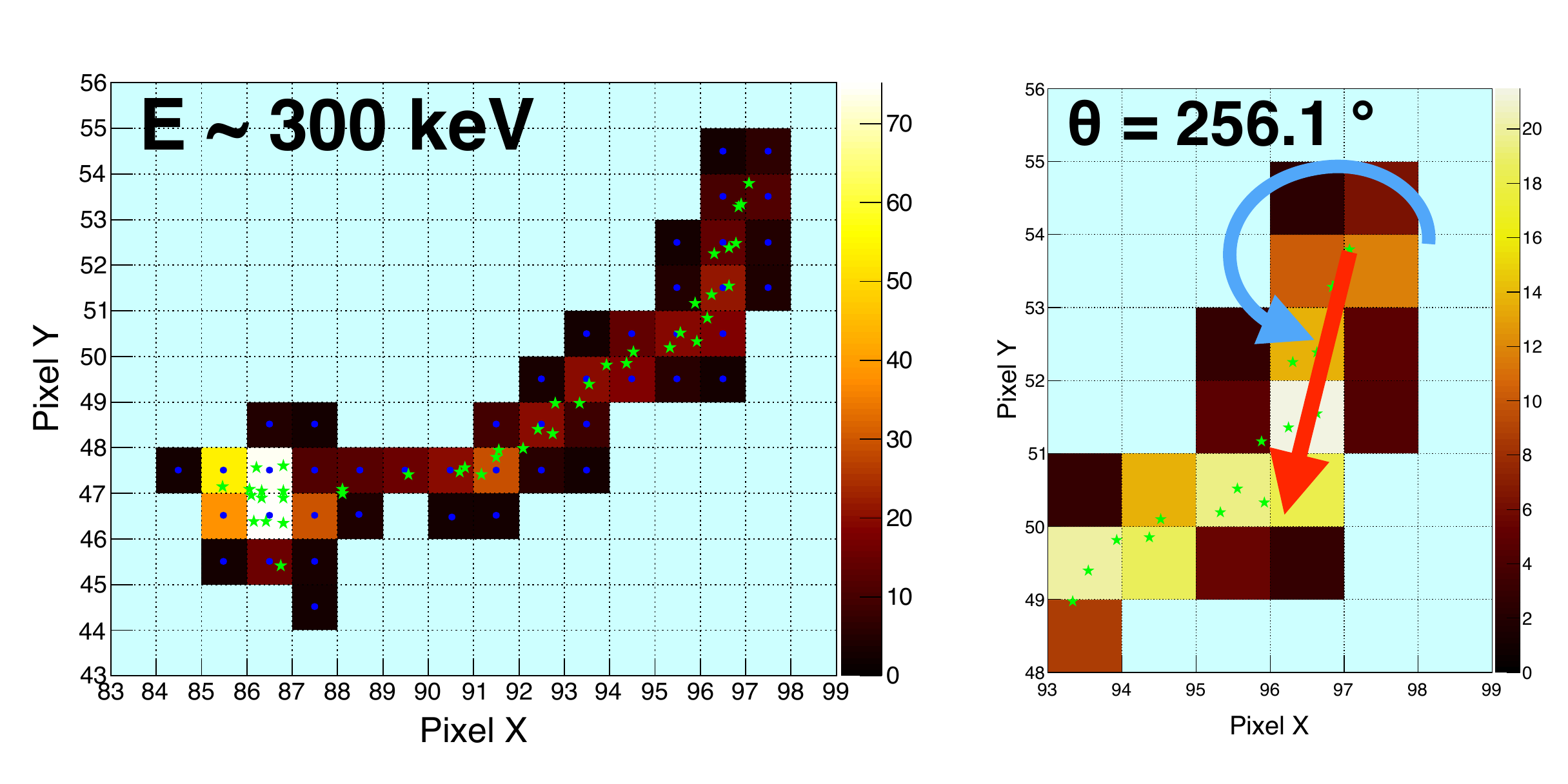}
        \caption{The reconstructed initial direction of an electron. (left) Blue points are the initial position of the nodes. Green points are the center of gravity of every pixels. (right) Red arrow is the reconstructed direction.}
    \label{fig:angle_cal} 
  \end{center}
\end{figure}

\begin{enumerate}
 \item{{\bf Making mathematical graph}}
 The center points of each pixels is regarded as vertices (the blue points in Figure~\ref{fig:angle_cal} ) of an graph. 
 The vertices are connected if their pixels are adjacent to each other and
 these connections are regarded as edges.
 Then these vertices and the edges form the graph.
 \item{{\bf Calculating the center of gravity of pixels}}
 Charges are shared with adjacent pixels thus their pixel values reflect the position information of the recoiled electron.
 Considering the charge sharing effect, we select each pixel and its adjacent pixels and calculate the center of gravity of these pixels 
 weighting with their pixel values and replace each node by the center of gravity.
 The blue points in Figure~\ref{fig:angle_cal} are the initial positions of the nodes
 and the green points are the replaced positions.
 After the replacement, the nodes trace the electron trajectory well.
 \item{\textbf{Finding longest \textit{shortest-path}}} 
 We select an arbitrary pair of the nodes and calculate its \textit{shortest-path} which connects the two nodes
 using Warshall-Floyd algorithm\citep{floyd:62}.
 This is a kind of the shortest path problem in graph theory.
 We determine the trajectory of the electron as the longest \textit{shortest-path} in all pairs.
  \item{{\bf Determining the Compton scattering position}} 
  The end points of the calculated path are the initial position and the stop one.
  We select each endpoint and its nearby $n_0$ nodes ($n_0$ are a parameter in this algorithm)
  and calculate the sum of their pixel values.
  The end point with the larger sum corresponds to the Bragg peak, thus
  the other is determined as the initial position ($\vec{x_0}$).
 \item{{\bf Applying principle component analysis}} 
 We select nearby $n_1$ nodes from the initial point 
 and apply the principle component analysis to these nodes and the initial point
 ($n_1$ is also a parameter).
 The first principle axis $\vec{v_e}$ is approximately parallel to the direction of the initial electron momentum.
 Thus the direction is $- \vec{v_e}$ or $+ \vec{v_e}$.
 We calculate the center of gravity of these points ($\vec{x_g}$) and
  select the direction which satisfies that the angle between $(\pm) \vec{v_e}$ and $\vec{x_g} - \vec{x_0}$ is less than 90$^\circ$.
\end{enumerate}

Figure~\ref{fig:angle_cal} shows the result applied this algorithm to the experimental data.
In this calculation, we set both $n_0$ and $n_1$ as 5.
The evaluation of the accuracy of this algorithm
and the comparison with other reconstruction algorithms (e.g. \citep{PLIMLEY:2011})
is under investigation.

%

\section{Conclusion}
We propose a new detector concept of Si-CMOS hybrid detector
towards electron tracking based Compton imaging with semiconductor detectors,
and developed two prototype detectors.
These prototypes are operated
and images of electron trajectories are produced successfully.
Electron energies are also measured from both the CMOS ROIC and the ASIC
and the energy resolutions (FWHM) are $4.1~\mathrm{keV}$ (CMOS) and $1.4~\mathrm{keV}$ (ASIC) at 59.5 keV.
Although there is a relatively large dead time ($\sim 15\%$) due to the interference noise currently,
ASIC events are identified with electron trajectories on CMOS images based on the ROI information.
Additionally, we developed an algorithm to measure the direction of the initial electron momentum based on graph theory, and confirmed that it works well by applying it to the experimental data.
The development of a semiconductor Compton camera using the second prototype of Si-CMOS hybrid detector is underway.

\section*{Acknowledgement}
The authors would like to thank Hamamatsu Photonics for supporting the development of the prototypes.
HY is financially supported by a Grant-in-Aid for Japan Society for the Promotion of Science (JSPS) Fellows (17J04145).
This work was supported by KAKENHI 16H02170, 24105007 and Advanced Measurement and Analysis program (JST-SENTAN Program, 2012-2014).

\section*{References}
\bibliography{mybibfile_v2}

\end{document}